\documentclass{article}
\usepackage{authblk}
\usepackage{hyperref}
\usepackage[utf8]{inputenc}
\usepackage{graphicx}
\usepackage{wrapfig}
\usepackage{amsmath,amsthm,pifont,amsfonts,MnSymbol}
\usepackage[a4paper, margin=4.5cm]{geometry}
\usepackage{tikz}
\begin{document}

\title{ Swarm consensus }
\author{Victor Grishchenko}
\author {Mikhail Patrakeev}
\affil{N.N. Krasovskii Institute of Mathematics and Mechanics,
Ekaterinburg, Russia}
\author{S. Q. Locke III}
\affil{unaffiliated}
\maketitle

\begin{abstract}
The strength of gnomes lies in their coordinated action.
Being small and subtle creatures themselves, the forest gnomes
can form large swarms acting as one giant creature.
This unusual defense strategy requires a lot of skill and
training. Directing a swarm is not an easy task!
Initially, gnomes used leader-based control algorithms, although
those have been proven to be vulnerable to abuse and failure.

After thorough research and study, gnomes developed their own
leaderless consensus algorithm based on very simple rules.
It is based on gossip in a network of a known diameter $d$.
One of the gnomes proposes a plan which then spreads gnome
to gnome.  If there is an agreement, gnomes act \emph{all at
once}.  If there are conflicting plans (an extreme rarity),
they try again. The resulting upper bound on the swarm's 
reaction time is its round-trip time $2dt$, where $t$ is the
command relay time. The original algorithm is non-Byzantine;
all gnomes must be sane and sober.

While working on the algorithm, gnomes discovered 
\emph{swarm time}, a sibling concept to L.~Lamport's logical time. 
That led to a Byzantine-ready version of the algorithm.
\end{abstract}

\theoremstyle{plain}
\newtheorem{teor}{Theorem}
\newtheorem{lemm}[teor]{Lemma}
\newtheorem{corr}[teor]{Corollary}
\newtheorem{prop}[teor]{Proposition}                                            
\newtheorem{rema}[teor]{Remark}
\newtheorem{algo}[teor]{Algorithm}
\theoremstyle{definition}
\newtheorem{defi}[teor]{Definition}
\newtheorem{term}[teor]{Terminology}
\newtheorem{nota}[teor]{Notation}
\newtheorem{conv}[teor]{Convention}
\theoremstyle{remark}
\newtheorem{exam}[teor]{Example}
\newtheorem{ques}[teor]{Question}

Running a swarm requires perfect coordination and consensus.
It has to adapt rapidly to the changing environment, as
well as its own changing form and composition.  As certain
incidents have shown, the centralized mode of coordination is
prone to abuse and failure.  Namely, the leader may prioritize
his own personal interest, while disappearance of the leader surely
disorganizes the swarm. Gnomes had to invent a better way!

They started by thinking their assumptions and limitations
through. First of all, gnomes can join a swarm at any time, then
fall off and rejoin at any rate.  Hence, any complex 
division of roles is impractical; especially, everything
election-based.  Also, a gnome's communication ability is
limited and the processing ability is even more limited. Any
sophisticated algorithms are definitely out of question. Most of
the gnomes can only repeat and relay commands, while also doing
their physical work.

\begin{wrapfigure}{r}{0.6\textwidth}
  \includegraphics[width=0.6\textwidth]{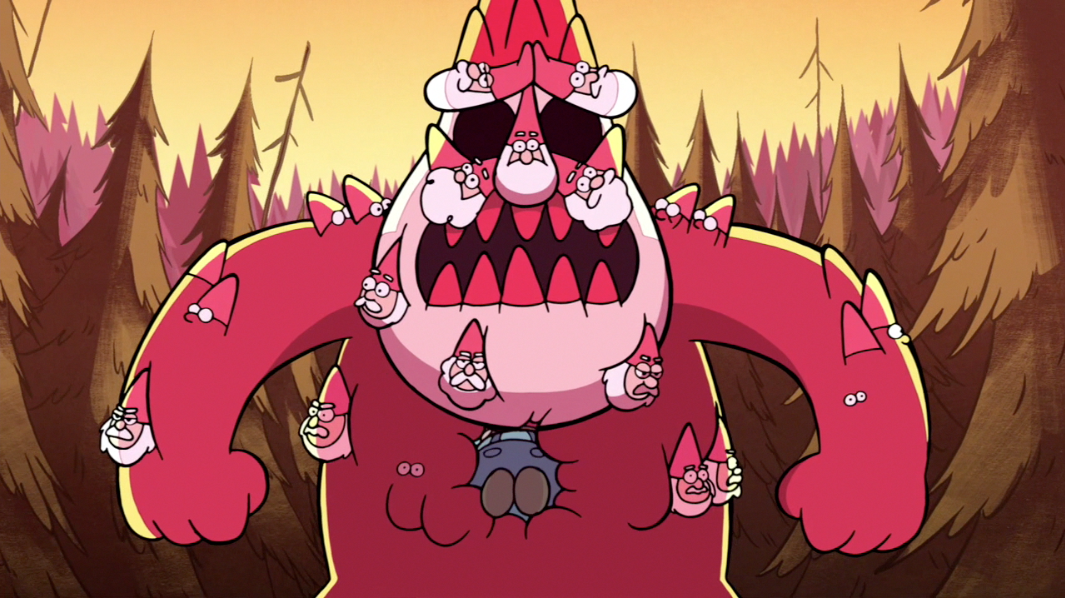} 
  \caption{An artist's impression of a gnome swarm. \copyright ~Disney} 
\end{wrapfigure}
Luckily, gnomes can relay messages very
efficiently, in uniform time $t$. While they did not want a
central leader anymore, they have pretty good ethics. So, they
may rely on the existence of naturally emerging leaders who will
propose necessary maneuvers at the right time. They rarely have
many proposers at once, but there is always someone. Finally, a
swarm requires that all gnomes act synchronously, without a time
lag. That was the most tricky requirement.

Gnomes studied the literature on databases and distributed
systems, as well as on swarm behavior of fishes, birds and
humans.  They studied 2 phase commit~\cite{2phase}, 3 phase 
commit~\cite{3phase}, Paxos~\cite{paxos} and Raft~\cite{raft}.
Sadly, all these algorithms focus on a consensus between
a small number of known participants.  Similarly, they imply the
existence of a leader giving orders unilaterally.  Even if the
leader is elected, that poses a significant communication
overhead and delay.  Gnomes can tolerate a temporary confusion,
but voting, counting and doing complex moves all at once is just
too difficult.  By the end of such an election the gnomes will
be laying on the ground, in the most optimistic case! Whom that
leader will command then?

Proof-of-work~\cite{bitcoin} (aka the Nakamoto consensus) was
laughed off. Gnomes will not burn their limited energy simply to
prove they are decent gnomes! Gnomes know each other, to
start with, so the problem itself is non-existant, not to
mention the price of the solution.  Similarly, the Byzantine
generals problem was found to be irrelevant at first.  Gnomes know
who their friends are.  What gnomes actually needed was a
non-Byzantine consensus algorithm for larger dynamic groups
where participants may come and go.  Also, their communication
ability is limited, so a gnome can only talk to his swarm
neighbors, not all the gnomes at once. Luckily, any corruption
of a message is very easy to detect when everybody is relaying
it and you can hear it from all sides.

The work led by late S.~Q.~Locke Sr resulted in a consensus
algorithm based on buddycast message relay in an open network.
Neither the full list of participants, nor their number is known
at any time.  The exact network topology is quite dynamic, so
better to consider it an unknown as well.  The only postulated
requirement is that the diameter of the network is bounded and
the upper bound $d$ is known. In other words, any gnome can reach any
other gnome through a neighbor chain of $d$ steps or less.  In a
swarm, that mostly occurs naturally.

\paragraph{The consensus algorithm} is simple: one gnome
proposes an action; other gnomes relay it if they consider it
reasonable. If all gnomes agree, they act all at once. The
tricky part is to know whether all gnomes agree on an action.
Preferably, quickly. Preferably, at once.  The exact algorithm
is as follows:

\begin{enumerate} 

\item   Gnome $p \in \mathbb{G}$ proposes an action to his swarm
        neighbors $n \in N(p)$, where $N(g)$ is the set of gnomes
        who can hear $g$, including himself.  Note that $h \in
        N(g) \iff g \in N(h)$.  We define a k-neighborhood of
        gnome $g$ as a set of gnomes being at most $k$ steps
        away: $N^{-1}(g):=\emptyset$, $N^0(g):=\{g\}$,     and
        recursively for $k \ge 0$, $N^{k+1}(g) := \bigcup_{h\in
        N^{k}(g)} N(h)$.  Note that $N^d(g) = \mathbb{G}$ where
        $d$ is the upper bound for the diameter of the swarm and
        $\mathbb{G}$ is all the swarm's gnomes. 

\item   Every gnome $g$ hearing an action proposed, relays it to his
        neighbors $N(g)$.  Note that gnomes have no difficulties relaying
        and listening to all the neighbors at once, and for each
        gnome, it takes exactly the same time to make an announce.
        For that reason, we speak of \emph{turns} of conversation.  

\item   Each gnome $g$ tracks the spread of the proposal in the
        following way: once all his neighbors $n \in N(g)$ say their
        $k$-neighborhood (or bigger) is aware of the proposal,
        $g$ realizes that his ($k+1$)-neighborhood is now aware
        and announces that on the next turn.  Recursively, that
        announce may trigger \emph{awareness neighborhood}
        increases for the neighbors.

\item   Let us define a function $\alpha$ that tracks the radius
        of that I-know-that-they-know neighborhood.  For gnomes
        unaware of the action, $\alpha_0(g):=-1$.  For
        the proposer, we set $\alpha_0(p):=0$.  Then
        recursively, 
        $\alpha_{t+1}(g) := 1 + \min\{\alpha_t(n) : n \in N(g)\}$
        if $\exists n \in N(g), \alpha_t(n) \ge 0$.
        Otherwise, for the unaware gnomes, $\alpha_{t+1}(g)$ is still $-1$.
        The awareness neighborhood
        itself is thus $A_t(g) := N^{\alpha_t(g)}(g)$.

\item   Once a gnome's $d$-neighborhood agrees on the proposal, he 
        realizes  that everyone now agrees. Then, he acts immediately! 
        $\alpha_t(g)=d \Rightarrow A_t(g) = N^d(g) = \mathbb{G}$.
        That completes a consensus round; gnomes become ready for
        new proposals.

\item   Once a gnome hears of two contradicting proposals at once, 
        he gets confused and announces that $\alpha_t(g) = -\infty$. 

\item   Confusion spreads as gnomes relay it. That precludes the
        $d$-neighborhood consensus.  The proposers have to think again and
        maybe retry proposing after a timeout, in the new round.

\end{enumerate}

The key finding is that gnomes reach $d$-neighborhood consensus all
at once, so they act all at once! That fact is completely
independent of the number of the gnomes in the swarm or the way
they are arranged. The only requirement is any two gnomes being
connected by a neighbor-to-neighbor chain of length $d$ or less.
That is the Schmebulock's consensus theorem, so let us prove it
for the humans. Unless noted otherwise, we assume no confusion 
happens.

\begin{lemm} The awareness neighborhood grows,
    $\alpha_{t+1}(g) \ge \alpha_t(g)$.
\end{lemm}
\begin{lemm}
    Once $\alpha_t(g) \ge d$, $g$ knows that all gnomes
    are aware of the proposal.
\end{lemm}
\begin{defi} 
    The swarm's \emph{bottom} is the gnomes who progressed the
    least in reaching the consensus: their awareness
    neighborhood radius is the smallest, 
    $b_t := \min \{ \alpha_t(g) : g \in \mathbb{G} \}$,
    $B_t := \{ g \in \mathbb{G}: \alpha_t(g) = b_t \}$.
\end{defi}

Suppose $r(p)$ is the maximum distance from the proposer $p$ to any
other gnome, $r(p) = \min \{ k : N^k(p) = \mathbb{G} \} $.
Then, $p$'s most remote gnomes are $R_p = \mathbb{G} - N^{r(p)-1}(p)$.
Note that at turn $r(p)-1$, the remote gnomes are not aware of the
proposal yet, $\forall g \in R_p, \alpha_{r(p)-1}(g)=-1$, while some their
neighbors just became aware. Next turn is $r(p)$ when all remote
gnomes become aware, while all their neighbors stay with $\alpha_{r(p)}(g)=0$.
The rest of the swarm will have $\alpha_{r(p)}(g)>0$
already. Following this dynamics, we can see that
\begin{lemm}\label{bottom}
    Initially, $b_{r(p)-1}=-1$ and $B_{r(p)-1}=R_p$.
    Later, for $t \ge r(p)-1$, 
    $b_{t+1} = b_t + 1$ and
    $B_{t+1} = \bigcup_{i \in B_t}N(i)$.
\end{lemm} 
In other words, $b_{r(p)+k}=k$ and all the bottom's neighbors 
also join the bottom.
Naturally, in another $d$ turns $b_{r(p)+d}=d$ and $B_{r(p)+d} =
\mathbb{G}$, so all gnomes act on turn $r(p)+d$ the
latest. Then, the question is: can any gnome act earlier than
turn $r(p)+d$? None can, but let us
suppose by contradiction that there is a gnome 
$h$ such that $\alpha_{r(p)+d-j}(h) \ge d$ for $j \ge 1$.
By applying the definition of $\alpha$ recursively we get:

\begin{lemm}
    If $\alpha_k(h) \ge l$, $0 \le i \le l$ and
    $e \in N^i(h)$, then $\alpha_{k-i}(e) \ge l - i$.
\end{lemm}

Then, if substituting $i, l$ for $d$, $k$ for $r(p)+d-j$, we get that
$\forall e \in N^d(h)$, $\alpha_{r(p)+d-j-d}(e) \ge d-d$, so
$\alpha_{r(p)-j}(e) \ge 0$, where $e$ might be in $R_p$ as
$N^i(h)=N^d(h)=\mathbb{G}$ 
which contradicts the fact that $\alpha_{r(p)-j}(e) \le
\alpha_{r(p)-1}(e) = -1$.

\begin{teor} \label{teor:sch}
    On turn $2d$ the latest, gnomes will become aware that they
    all agree, all at once. 
    $\exists t \le 2d: \forall g \in \mathbb{G},~
     \alpha_t(g) = d$ and $\forall t' < t: \alpha_{t'}(g) < d$.
    \qed
\end{teor}

\begin{figure}[h] 
	\centering
  \includegraphics[width=0.95\textwidth]{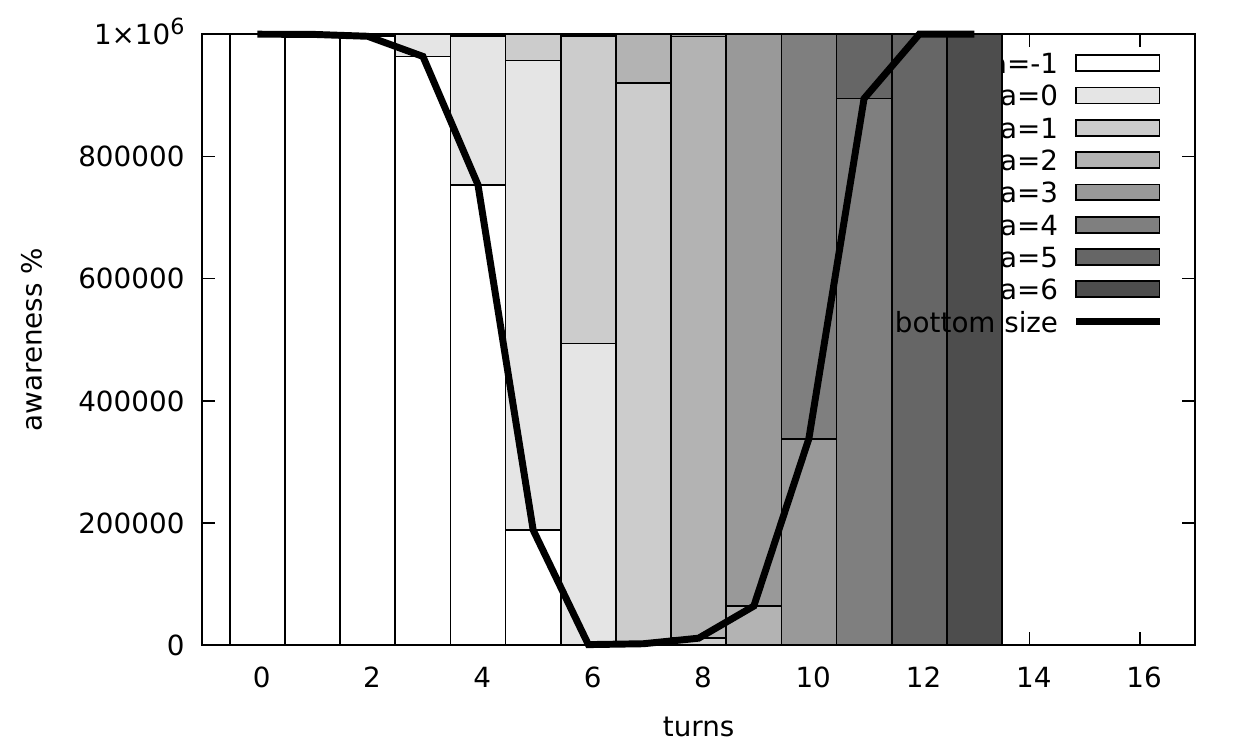} 
  \caption{Simulation: one-million gnome swarm synchronizes
    in 14 turns. Bars show the percentage of gnomes progressed
    to a certain awareness neighborhood radius $\alpha_t(g)$. 
    The line shows the number of bottom gnomes.} 
    \label{mln}
\end{figure}

A gnome swarm rarely reaches a thousand members, but we made a
computer simulation for a million-strong swarm and the algorithm
works well in that case, see Fig.~\ref{mln}. For many common graph
topologies, the diameter is logarithmic to the size.
For the social graph of humanity, $d$
is believed to be $6$ (``six degrees of separation'').
Hence, the swarm reaction time can be small for very large
graphs. 
Also, the number of messages a node has to process
is linear to the number of its edges. In other words, the
algorithm is very scalable. Many consensus algorithms require
every node to talk to or at least to be aware of every other node. That
makes the number of messages quadratic to the graph size $O(N^2)$.
In our case, a feasible upper bound is $O(N e \log N)$ messages,
where $e$ is the number of edges per node.

\paragraph{Swarm time} is a wonderful concept which is both
similar to and the exact opposite of L.~Lamport's \emph{logical
time}~\cite{logical}.  Remember that we defined $\alpha_t(g)$ as
the radius of a gnome's I-know-that-they-know neighborhood.  After
$2d$ turns, the swarm synchronizes so $\forall g,h \in
\mathbb{G}: \alpha_{2d+k}(g)=\alpha_{2d+k}(h)$. Still, nothing
prevents the gnomes from counting indefinitely. Beyond $d$
steps, the notion of \emph{neighborhood} has no meaning:
$N^{d+k}(g)=N^{d+l}(h)=\mathbb{G}$.  Surprisingly, we can also
see $\alpha_t$ as a time metric as it is monotonous for each
gnome and roughly synchronous for the swarm.  The Schmebulock's
algorithms lets a swarm create a shared clock!  In the simplest
and most popular case, Lamport's logical time is defined as a
\emph{maximum} of incoming time values plus 1. The swarm time is
a \emph{minimum} of incoming time values plus 1.  The mission of
both metrics is exactly the same: establish a shared clock based
solely on message passing between distributed processes.  The
way of timekeeping is a cornerstone of any distributed
architecture.  The BitCoin paper~\cite{bitcoin} describes most of its machinery
as a way to ``implement a distributed timestamp server on a
peer-to-peer basis''. The Google Spanner paper refers to its
satellite-based TrueTime system as ``the key
enabler''~\cite{spanner}.  Authors believe that swarm time is
the best fit for massively distributed self-synchronizing
swarmed systems that we yet have to build.

\paragraph{Real-world considerations} for the algorithm are
manifold. Three main concerns are node churn, varying
transmission times, faulty and Byzantine behavior. Indeed, gnomes may fall
off the swarm, some gnomes might be slower than others, and
finally, some may be unsober. To address that, we make three
separate generalizations to the algorithm.

To account for node churn, we add two rules. First, a newly
joining gnome must be ignored till the start of the next round.
Second, the diameter upper bound $d$ must hold. Re-assessing $d$
in a moving swarm is difficult, so gnomes must keep the actual
diameter under that pre-agreed bound at all times.

To account for the varying speed of message passing, we have to
separate the physical time $\tau$ measured in seconds from the
logical turns of the message exchange.  As before, $\tau=0$ is
the moment of the proposer's initial announce.  Assuming every
gnome is able to convey his state change in $\tau_{\max}$
seconds or less, the physical swarm diameter is $D \le
d\tau_{\max}$.  Then, $\alpha_{\tau}(g)$ is defined as 1 plus
the minimum of $\alpha_{\tau'}(n)$ so far received from
$n \in N(g)$.  
This change does not affect the dynamics of
the algorithm much as \emph{min} is not sensitive to the order of
arguments. Hence, the relative speed of updates does not change
much compared to the discrete case. 

Finally, to address Byzantine faults we would need both swarm time
and a multiphase consensus strategy. We define a gnome's \emph{consensus
phase} as his degree of knowledge about the swarm's knowledge.
Namely, a gnome has phase 0 if he is aware of the proposal.
A gnome has phase $k+1$, 
iff he is aware of the proposal and knows
that all other gnomes reached phase $k$ on that proposal.
Note that a gnome reaching phase $k$ also has all the earlier
phases. 
Let us formulate the Schmebulock's theorem with all those
generalizations in mind.  
\begin{defi}
The swarm's \emph{bottom time} is the minimum swarm time at
a given moment $\tau$ of physical time, 
    $\mu_\tau := \min \{ \alpha_\tau(g) : g \in \mathbb{G} \}$.
\end{defi}
\begin{lemm}
    Similarly to Lemma~\ref{bottom}, 
    if $\mu_{\tau} \ge 0$, then $\mu_{\tau+\tau_{\max}} \ge
    \mu_{\tau} + 1$.
\end{lemm}
\begin{corr}
    $\mu_\tau \ge \lfloor \frac{\tau}{\tau_{\max}} \rfloor - d$.
\end{corr}
\begin{lemm}
    For any two gnomes, the swarm time differs at most by $d$.
\end{lemm}
\begin{lemm}
    At time $\tau$, $g, h \in \mathbb{G}$, then $g$ knows that 
    $\alpha_\tau(h) \ge \alpha_\tau(g) - d$.
\end{lemm}
\begin{teor}
    At time $\tau \ge 0$, each gnome has consensus phase 
    $\lfloor \frac{\tau}{d\tau_{\max}}\rfloor - 1$.
\end{teor}

\paragraph{Byzantine attacks} can be a serious issue for a swarm once
some gnomes abuse certain mushrooms and consequently become
unsober and unreasonable. As a first mitigation, there is a list of
simple sanity checks swarm neighbors use to expel violators:

\begin{enumerate}
    \item a gnome can not backtrack by announcing a smaller
        neighborhood number than before, $\alpha_{t+1}(g) \ge \alpha_t(g)$,
    \item a gnome can not stay at the same neighborhood number for
        more than 2 turns (unless the gnome is unaware of any
        new actions), $\alpha_{t+2}(g) > \alpha_t(g)$,
    \item a neighbor can not announce a number greater than the
        number we announced to him, plus one
        $n \in N(g)$ then $\alpha_{t+1}(n) \le \alpha_t(g) + 1$,
    \item a gnome can not announce a different action unless the
        previous action was completed or the confusion timeout
        has passed,
    \item a gnome can not keep progressing after being told that
        there is a conflicting proposal (must become confused).
\end{enumerate}

\begin{figure}[h!]
\usetikzlibrary{arrows}
\center
\begin{tikzpicture}

\draw[style=help lines] (0,0) grid (12,4);
\node[below,color=gray](p0) at (2,4) {phase 0};
\node[below,color=gray](p0) at (6,4) {phase 1};
\node[below,color=gray](p0) at (10,4) {phase 2};
\draw[ultra thick,color=lightgray,->](0,4) -- node[sloped,above]{turns} (12,4);
\draw[ultra thick,color=lightgray,->](0,4) -- node[sloped,below]{swarm} (0,0);

\node[left] (p) at (0,4) {$p$};
\node[left] (j) at (0,0) {$j$};
\node[below] (o) at (0,0) {$0$};
\node[below] (d) at (4,0) {$d$};
\node[below] (d2) at (8,0) {$2d$};

\draw[thick,-angle 45](0,4) -- node[sloped,below,font=\small]{proposal spreads} (4,0) ;
\draw[thick,-angle 45] (4,0) -- (8,4);
\draw[thick,-angle 45]((8,4)  -- (12,0);

\draw[very thick,dotted] (4,4) --node[sloped,below,font=\tiny]{all know} (4,0);
\draw[very thick,dotted] (8,4) -- node[sloped,below,font=\tiny]{all know: all know, none confused} (8,0);
\draw[very thick,dotted] (12,4) -- node[sloped,below,font=\tiny]{all know: all know all know} (12,0);

\draw[color=red,-angle 45] (2,0) -- (4,2);
\node[below,color=red] (i) at (2,0) {$t_{c}$};
\draw[color=red,dashed] (4,2) -- node[sloped,above,font=\small]{confused} (4,0);

\draw[color=red,-angle 45] (6,0) -- (8,2);
\node[below,color=red] (i) at (6,0) {$t_{f}$};
\draw[color=red,dashed] (8,2) -- node[sloped,above,font=\small]{fooled} (8,0);

\draw[color=red,-angle 45] (10,0) -- (12,2);
\node[below,color=red] (i) at (10,0) {$t_{t}$};
\draw[color=red,dashed] (12,2) -- node[sloped,above,font=\small]{tricked} (12,0);

\draw[thick,color=orange,dashed,angle 45-] (2,0) -- node[pos=0.25,sloped,below,font=\tiny]{backdating} (6,0);

\end{tikzpicture}

\caption{Byzantine sabotage: a competing proposal is injected
    under different phases of consensus. Propagation of the valid proposal
    is the solid black line, the malicious proposal is red. 
    The drawing is effectively a Minkowski diagram.}
    \label{sabotage}
\end{figure}
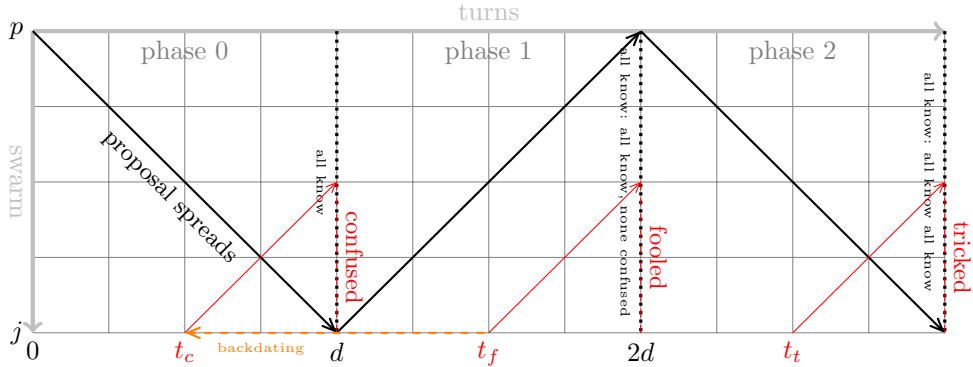

But, it becomes much worse if the unsober gnome feels ironic and
starts to sabotage the consensus in a smart way. Various
scenarios of sabotage are shown on Fig.~\ref{sabotage}.  There
are three phases of consensus pictured. After phase 0, all
gnomes know the proposal. After phase 1, all gnomes know
that all other gnomes know. After phase 2, gnomes know that all
know that all know.
Note that any event ripples through the swarm in $d$ turns.

The unsober gnome $j$ (the \emph{joker}) can inject a
competing proposal on turn $t_c$ during phase 0 to
\emph{confuse} the swarm and prevent $p$'s proposal from
achieving consensus.  Alternatively, he can inject on turn $t_f$
during phase 1 to \emph{fool} some part of the gnomes into
believing there are confused gnomes. It is forbidden to make
proposals in phase 1, but the joker may say his \emph{silly friend} is
confused. That would look plausible as normal confusion spreads
in phases 0 and 1.  Finally, a joker may inject a competing
proposal in phase 2 to \emph{trick} some gnomes into believing
he is fooled.  In each case, the swarm would divide into parts:
agreeing and confused, acting and fooled, non-tricked and
tricked.

One way to handle competing proposals in blockchain
architectures is to use some lottery to limit the
ability to propose. That might be a ``happy hash'' in
proof-of-work architectures or a ``happy second'' in some
proof-of-stakes. The idea of lottery is in serious conflict with
the efficiency of a swarm. What if the happy ticket falls to the
sleepy gnome, not to the wise one? 

More sophisticated techniques depend on swarm time being tracked
continuously across rounds. That is not too difficult.  Just
$2d$ turns after the start, gnomes synchronize and the swarm
sounds like it is buzzing a rhythmic tune. That puts all events
on a shared time/space grid, very much like Fig.~\ref{sabotage}
shows. Turn numbers can now be compared between different proposals.

Then, it is possible to apply a technique reminiscent of the
Lamport's \emph{arbitrary total order}~\cite{logical}. Except,
this one is based on the swarm time, not logical time. Once
proposals can be ranked by their creation time, any later
injections can be ignored. Ties are resolved based on the
proposer's \emph{rank}. A gnome's rank is not a static value, but the
details are irrelevant here. It is sufficient to say that gnomes
can uniformly choose between two competing proposals.  That
leaves one opportunity though: the joker may backdate his bogus
proposal. That can be done in the phase 0 only. If the proposal
is backdated more than $d$ turns, gnomes would realize: they
should have heard of it before but they did not. Such a joker
would be thrown out immediately. Similarly, a backdated proposal
may never reach consensus: on its nominal turn $2d$, its actual
turn would be less than that, so the awareness
neighborhood size may be too small $\alpha_{2d}(g) < d$.  The
bogus proposal would not be acted on, but it can outcompete
the valid proposal while propagating. 

That leads us to the last anti-Byzantine technique gnomes call a
``merry swarm''. They mostly do it for fun to see how long it
can hold while more and more gnomes misbehave. In a merry swarm,
gnomes let each proposal spread and act on the one that reaches
$\alpha_t(g)=d$ sooner.  If there is a tie, they resolve it in a
uniform arbitrary way.

\paragraph{  As a conclusion,  } we can only praise the
ingenuity of the gnomes.
Differently from the past consensus algorithms preoccupied with
leader elections and majorities, the Schmebulock's algorithm can
reach (non-Byzantine) consensus in an open network of arbitrary
topology, where each gnome is only aware of his immediate
neighbors and can only communicate with them.  Neither the total
number of gnomes nor the exact topology of the network are known
to the participants.  Nevertheless, gnomes are perfectly able to
synchronize their behavior and act all at once!

\bibliographystyle{plain}
\bibliography{schmebulock}{}

\appendix

\section{An example}

Here we show a consensus progression in a really small swarm
of 10 gnomes of unusually large diameter 5.

\includegraphics[width=0.5\textwidth]{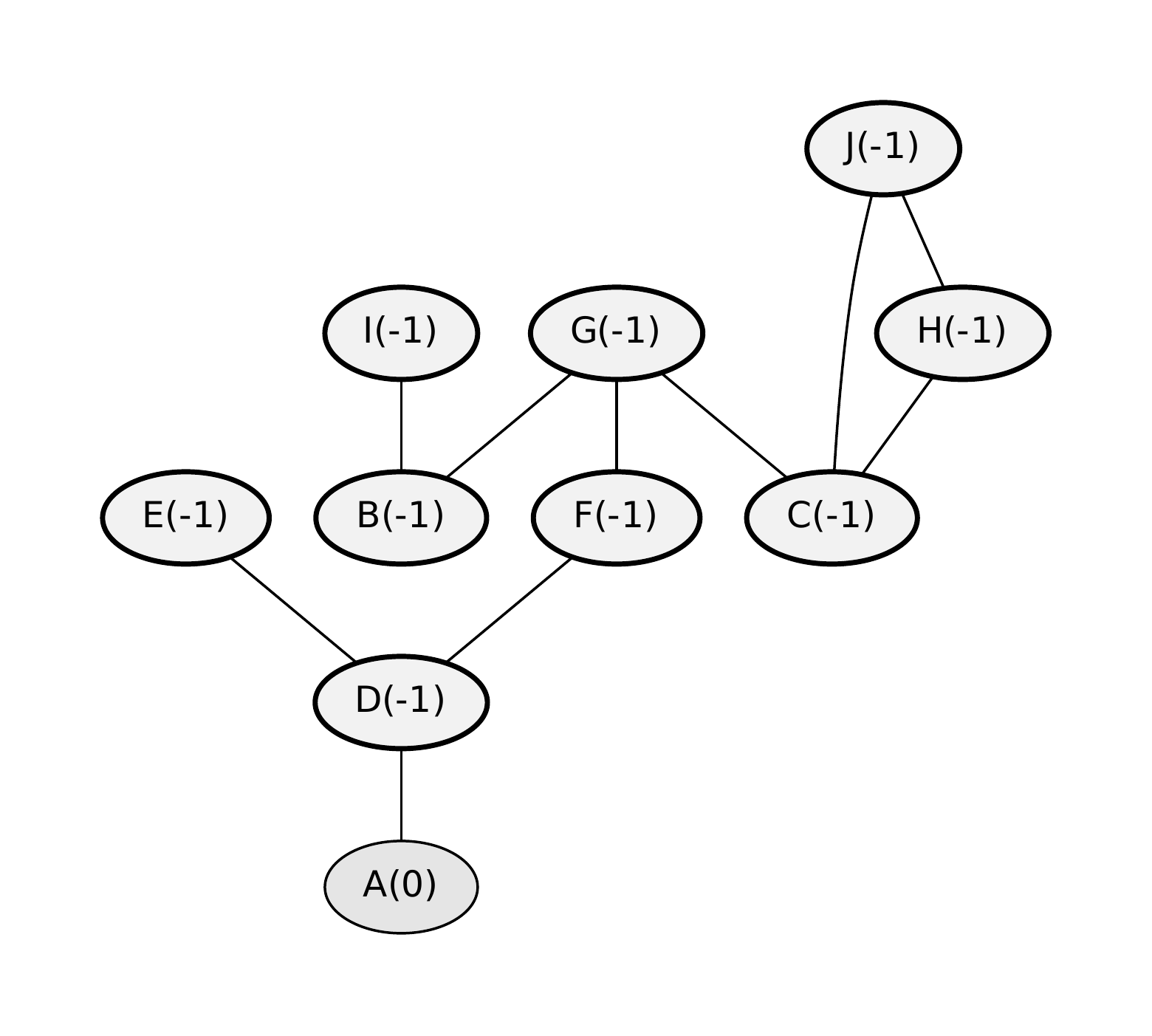}
\includegraphics[width=0.5\textwidth]{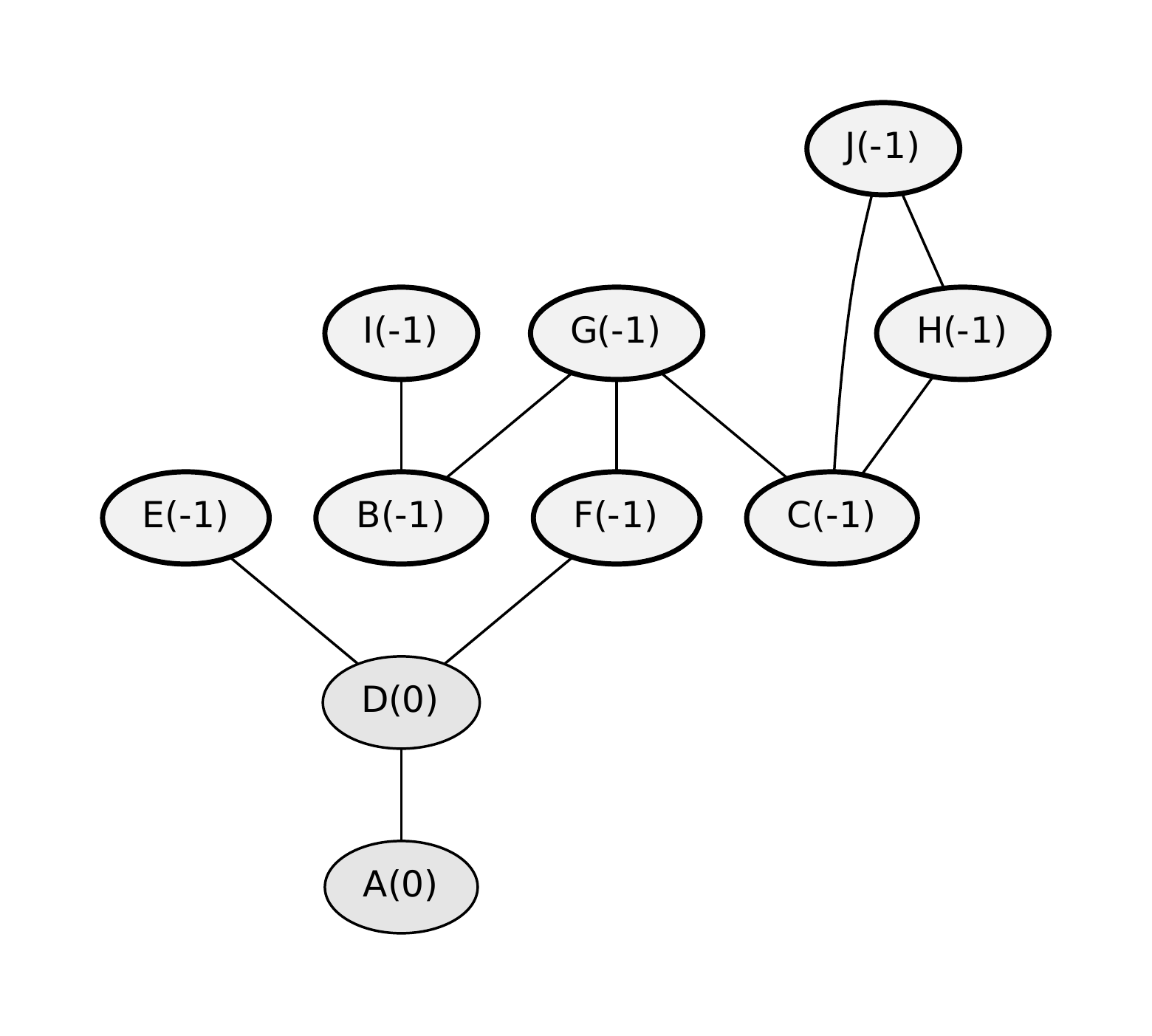}
\\~
\includegraphics[width=0.5\textwidth]{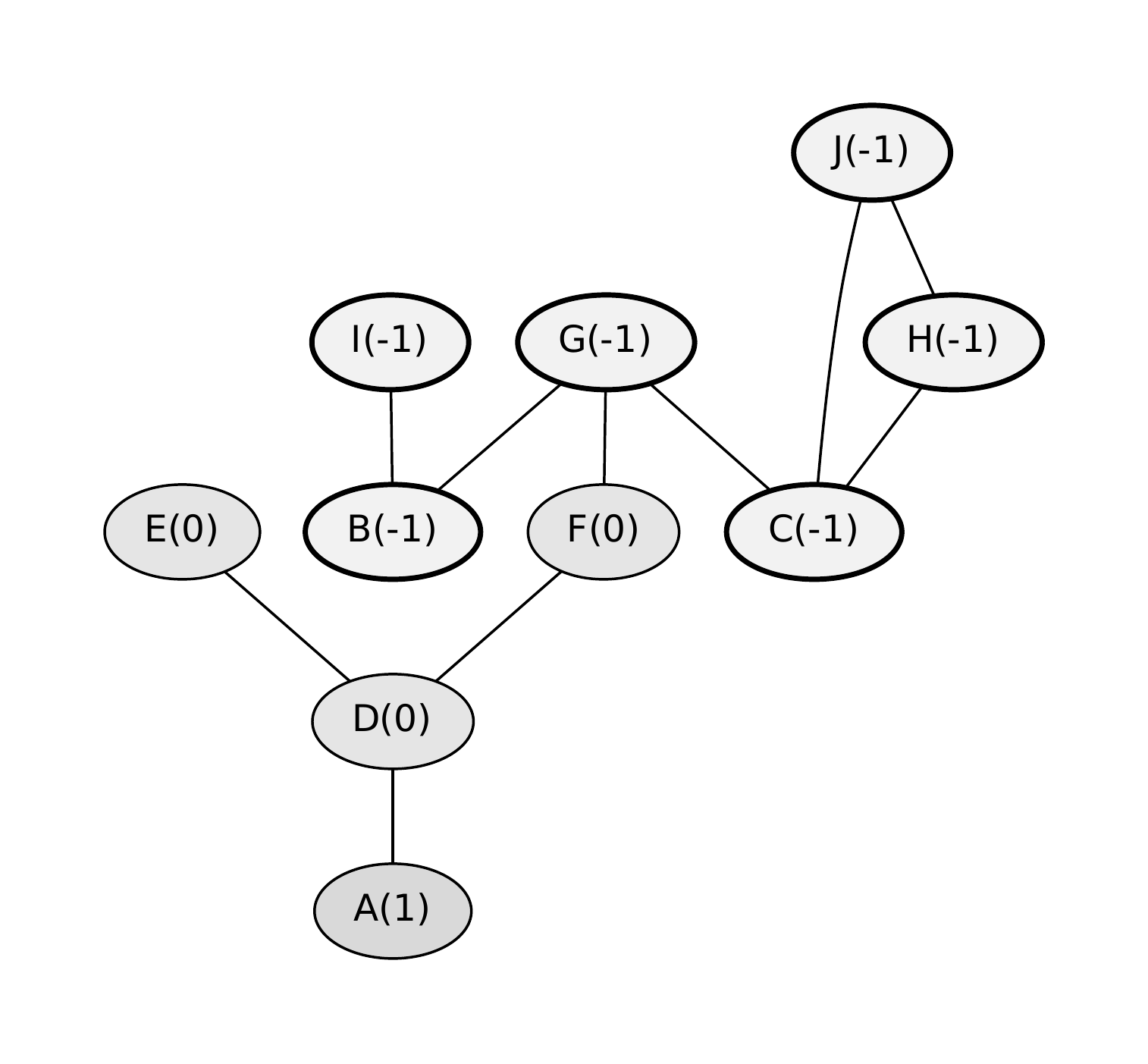}
\includegraphics[width=0.5\textwidth]{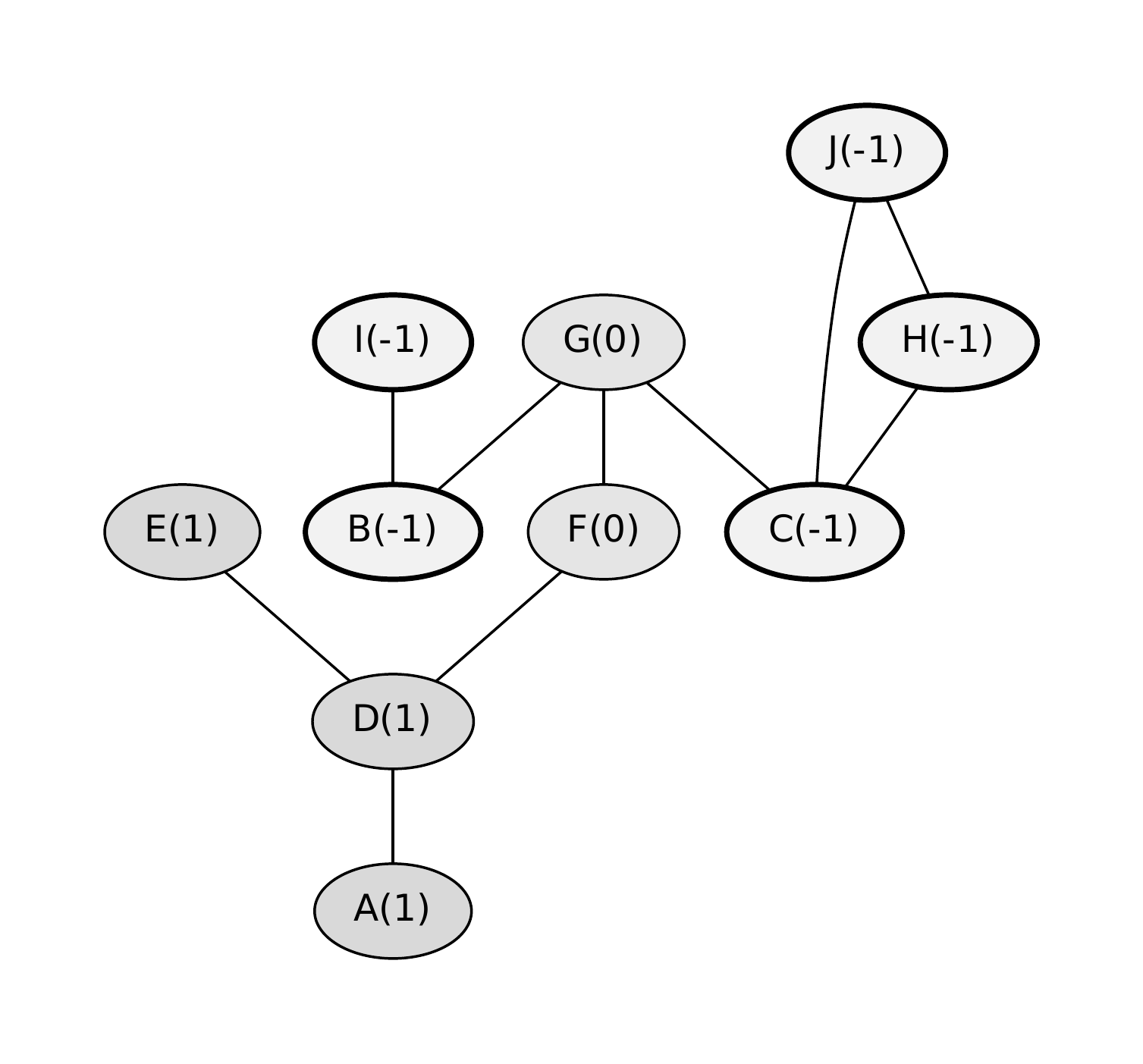}
\\~
\includegraphics[width=0.5\textwidth]{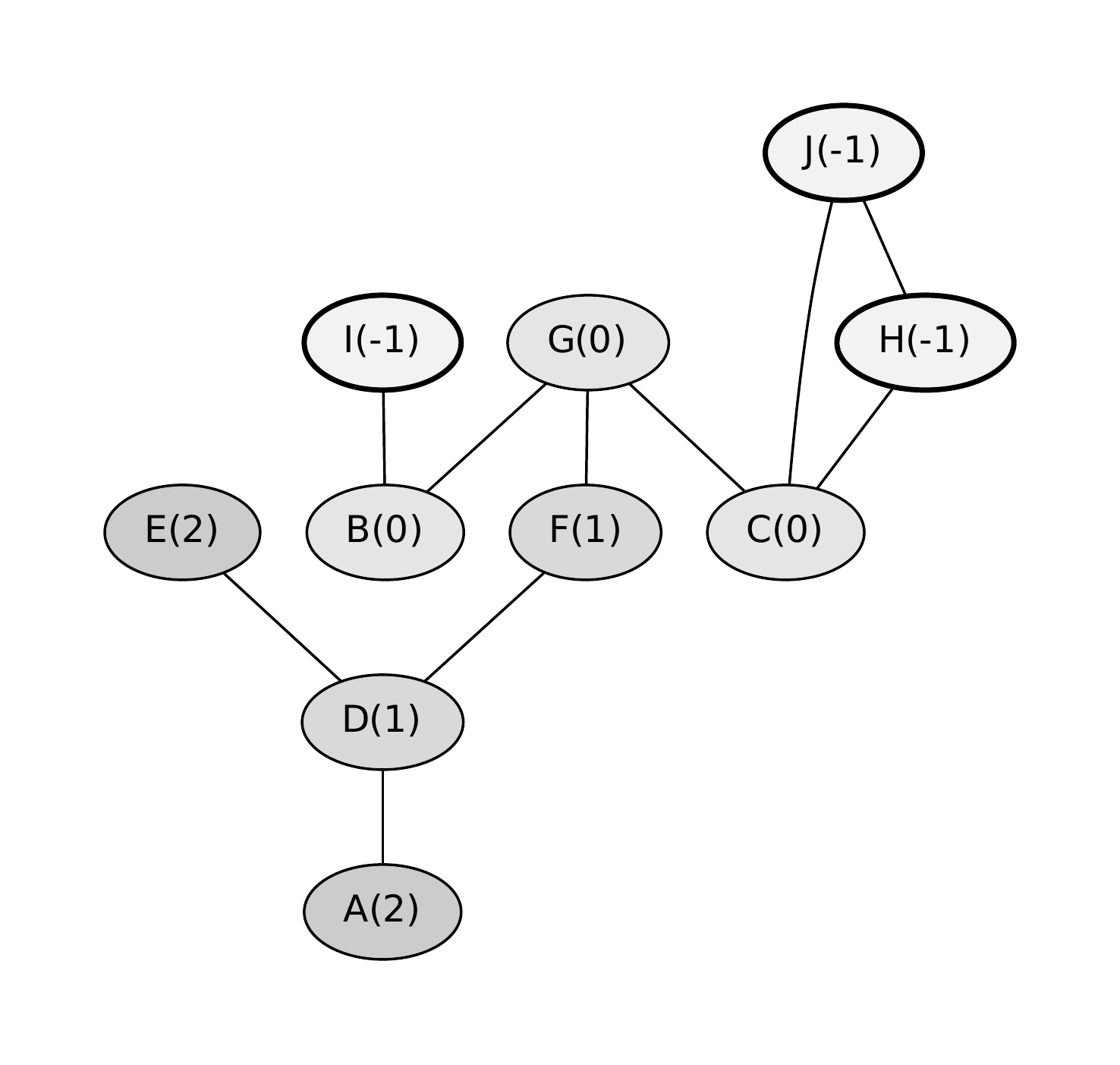}
\includegraphics[width=0.5\textwidth]{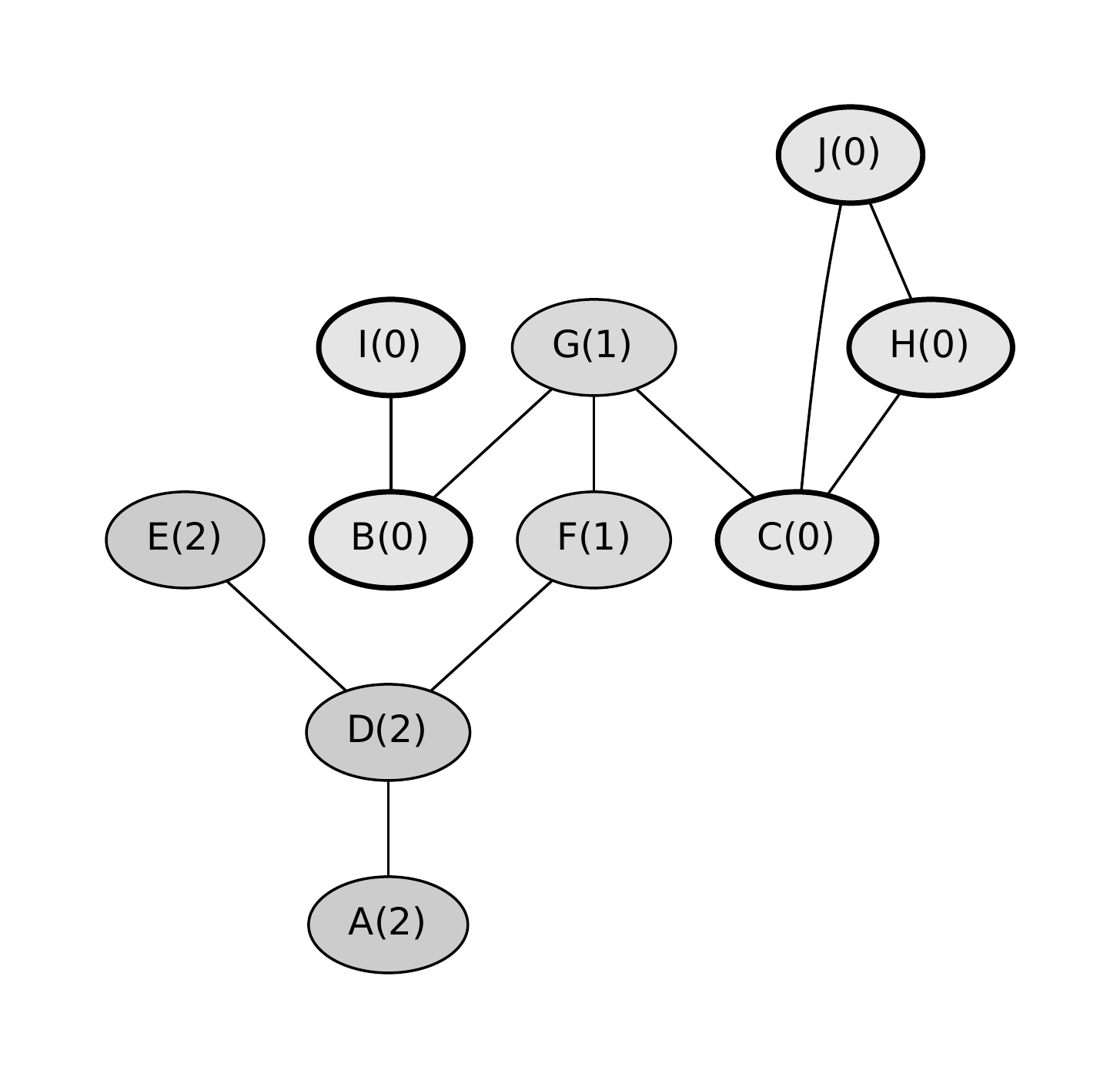}
\\~
\includegraphics[width=0.5\textwidth]{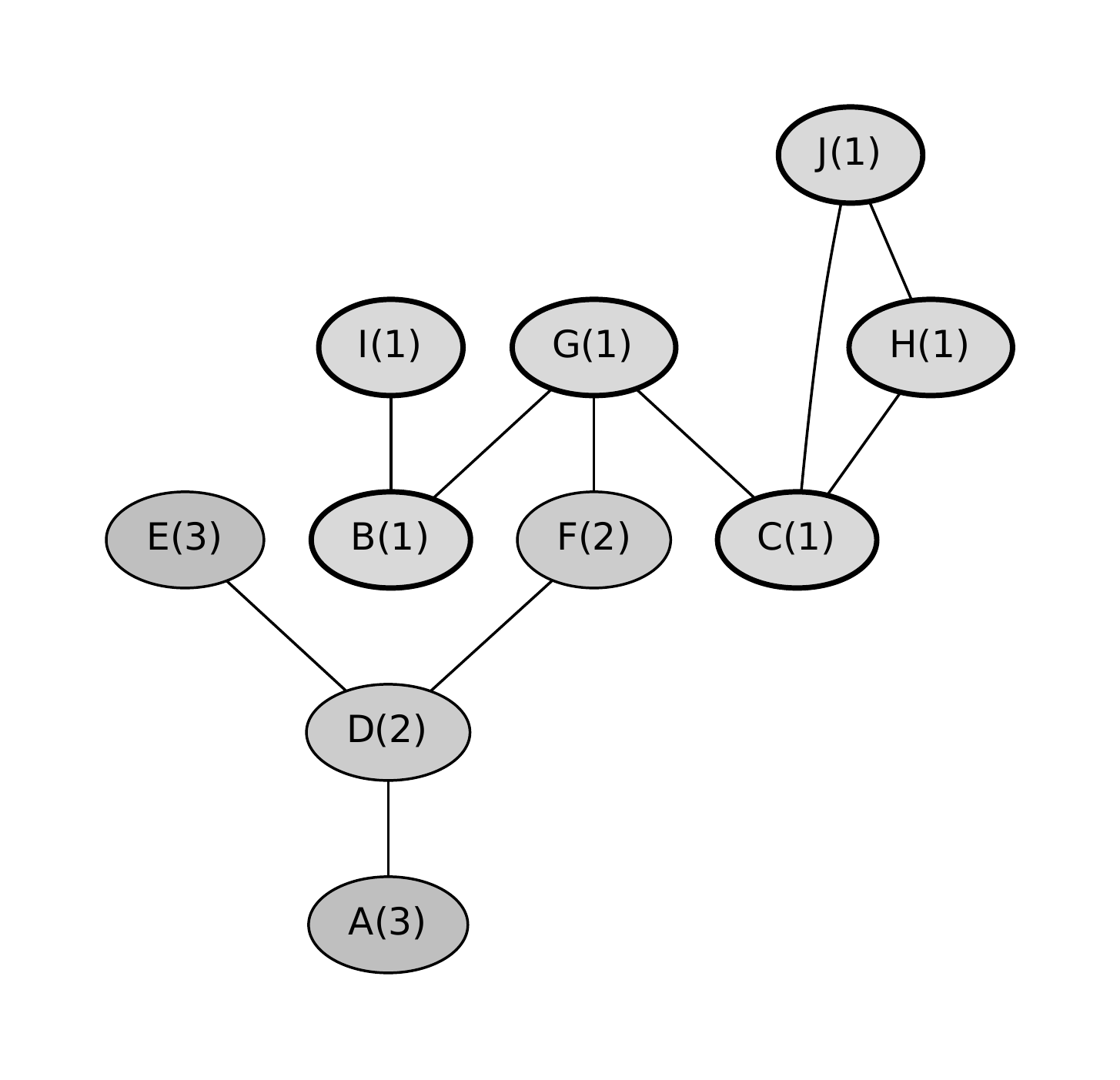}
\includegraphics[width=0.5\textwidth]{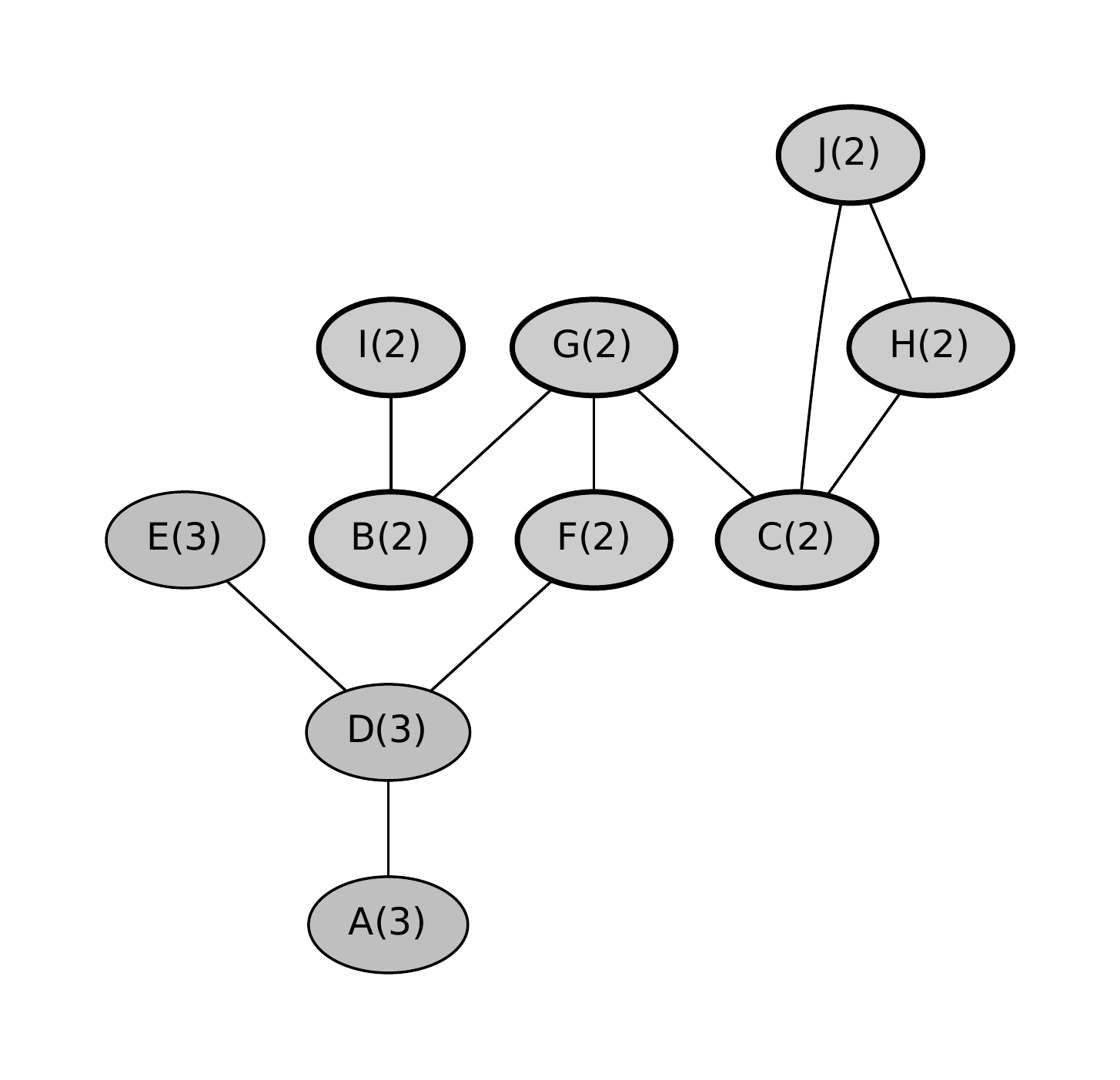}
\\~
\includegraphics[width=0.5\textwidth]{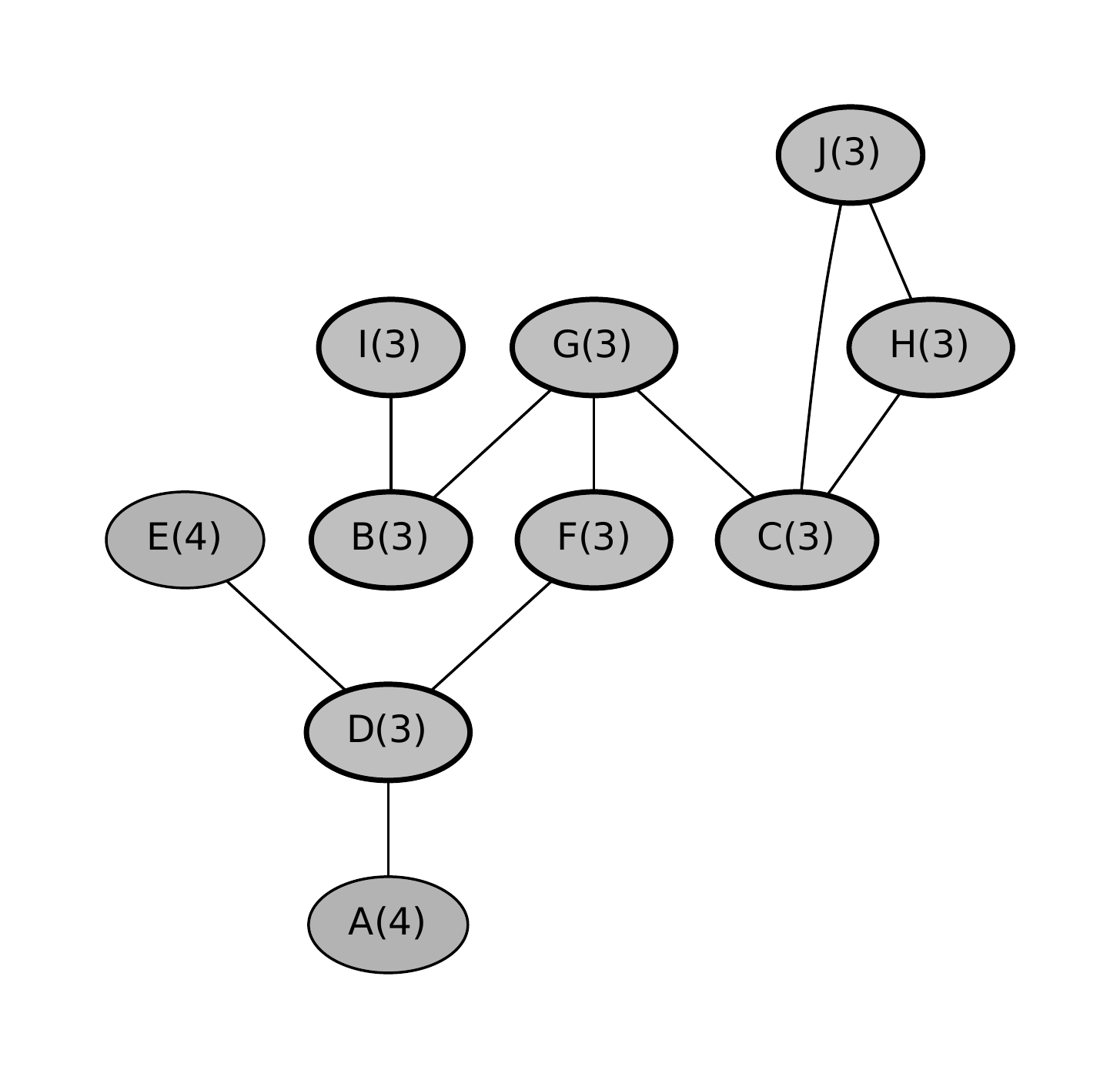}
\includegraphics[width=0.5\textwidth]{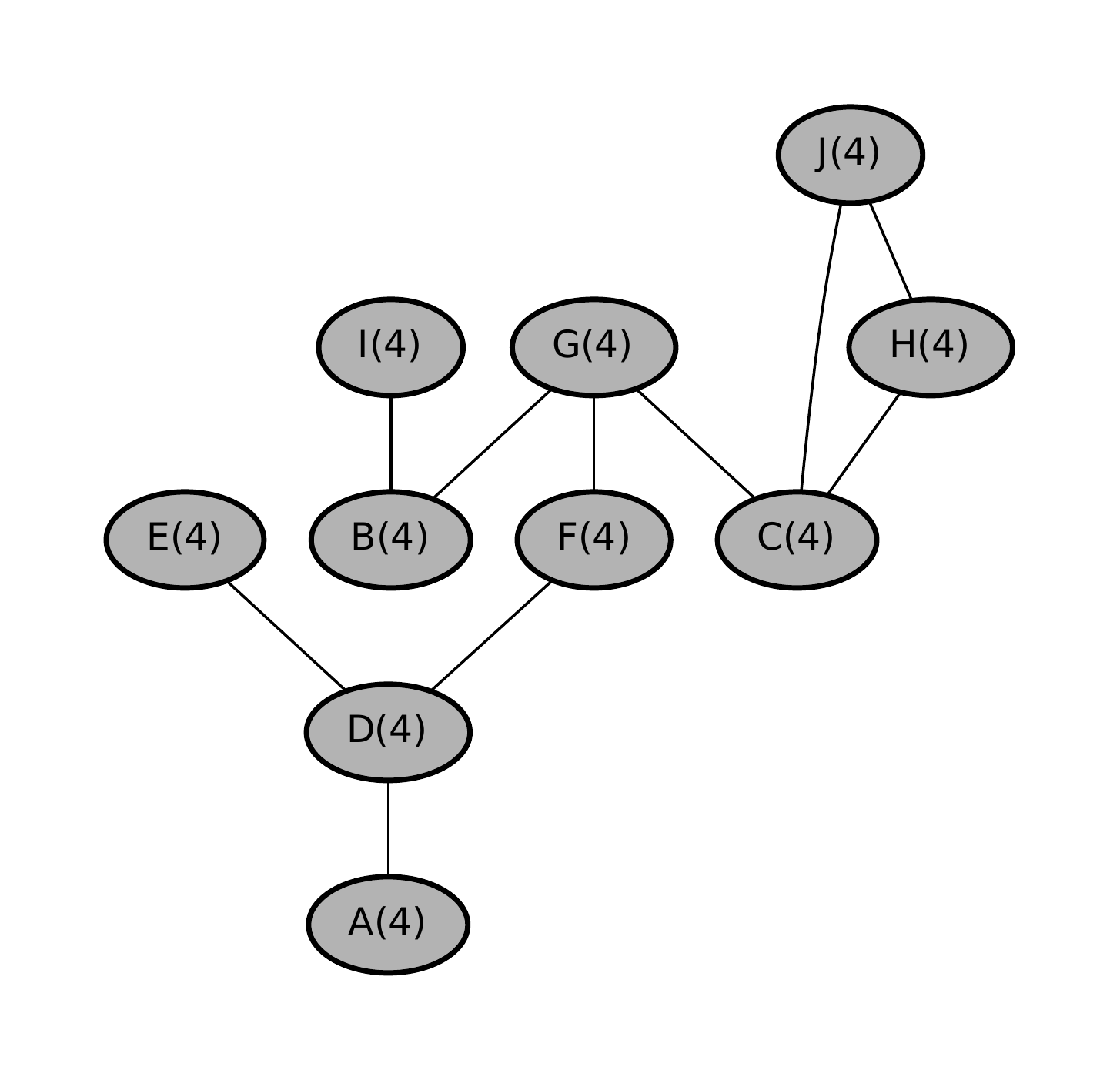}
\\~
\includegraphics[width=0.5\textwidth]{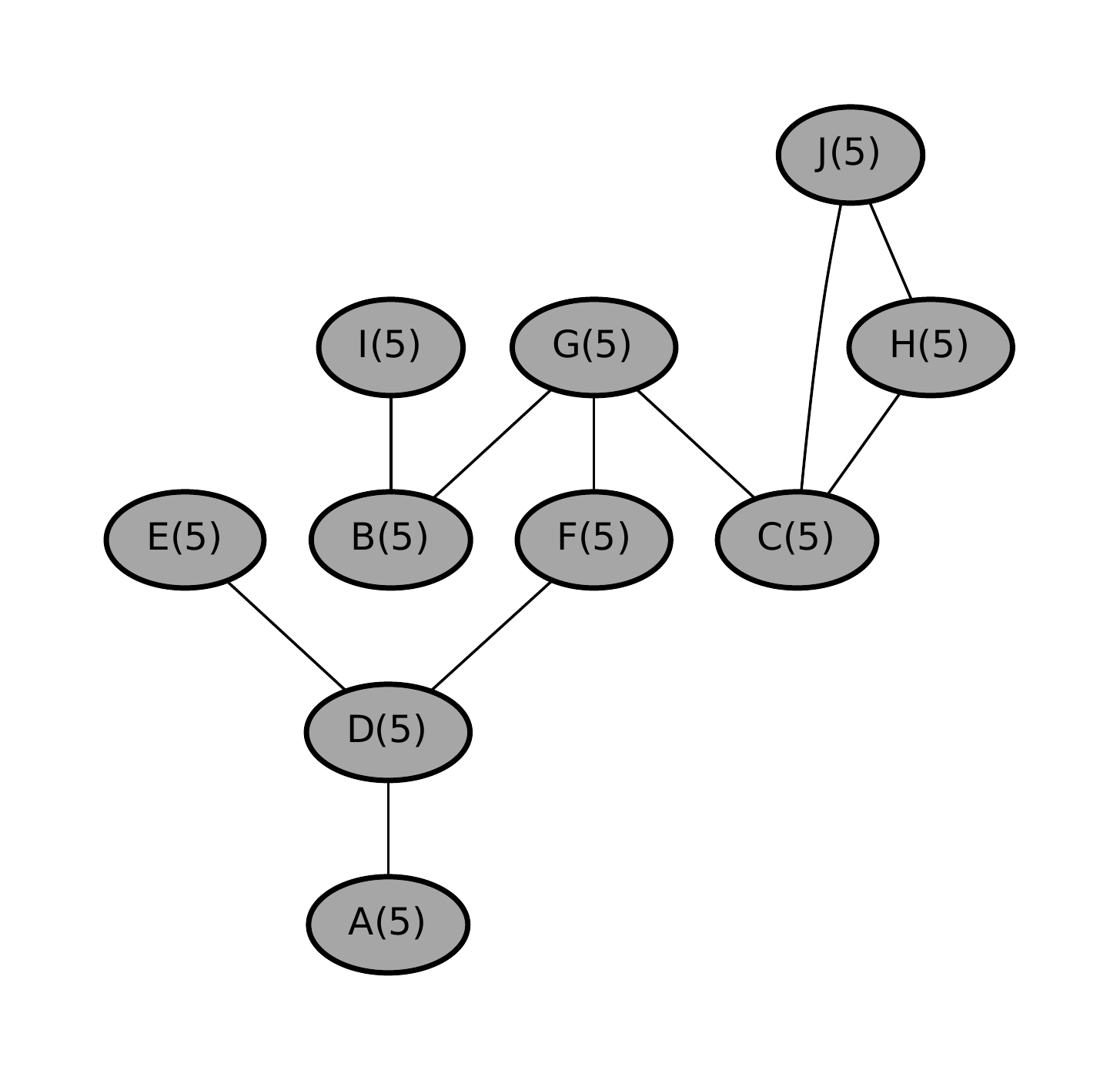}
\ldots et cetera


\end{document}